\documentclass
[showkeys,showpacs,preprintnumbers,nofootinbib]{revtex4} 
\usepackage{amsmath,amsfonts,amssymb,amscd,amsxtra,amsthm}
\usepackage{graphicx}
\usepackage{bm}
\begin{document}   
\preprint{CYCU-HEP-09-19}
\title{Chiral magnetic effect at low temperature}      
\author{Seung-il Nam}
\email{sinam@cycu.edu.tw}
\affiliation{Department of Physics, Chung-Yuan Christian University (CYCU), 
Chung-Li 32023, Taiwan} 
\date{\today}
\begin{abstract}  
We investigate the chiral magnetic effect (CME) under a strong magnetic field $\bm{B}=B_{0}\hat{x}_{3}$ at low temperature $T\lesssim T^{\chi}_{c}$. For this purpose, we employ the instanton vacuum configuration with the finite instanton-number fluctuation $\Delta$, which relates to the nontrivial topological charge $Q_{\mathrm{t}}$. We compute the vacuum expectation values of the local chiral density $\langle \rho_{\chi}\rangle$, chiral charge density $\langle n_{\chi}\rangle $ and induced electromagnetic current $\langle j_{\mu}\rangle$, which signal the CME, as functions of $T$ and $B_{0}$. We observed that the longitudinal EM current is much larger than the transverse one, $|j_{\perp}/j_{\parallel}|\propto Q_{\mathrm{t}}$, and the $\langle n_{\chi}\rangle $ equals to the $|\langle j_{3,4}\rangle|$. It also turns out that the CME becomes insensitive to the magnetic field as $T$ increases, according to the decreasing instanton, {\it i.e.}  tunneling effect. Within our framework, the instanton contribution to the CME becomes almost negligible beyond $T\approx300$ MeV.
\end{abstract} 
\maketitle
\section{Introduction}
Investigations on the nontrivial QCD vacuum structure have been one of the most important and dedicated subjects in modern particle and hadron physics for several decades. We note that the QCD ground state consists of infinitely degenerated vacua ($\theta$-vacuum), and each degenerated vacuum can be enumerated by its own integer topological number, {\it i.e.}  the Chern-Simon number $n_{\mathrm{CS}}$. According to the Atiyah-Singer theorem, two different vacua can be related to each other by a nontrivial topological charge $Q_{\mathrm{t}}=n_{\mathrm{CS}}(t=\infty)-n_{\mathrm{CS}}(t=-\infty)$, which stands for $\theta\ne0$ and signals the breakdown of the $P$- and $CP$-invariance of the vacuum. Since each vacuum is separated by a certain potential barrier, a nontrivial $Q_{\mathrm{t}}$ means quantum mechanical tunneling or passing over the barrier with enough energy. These novel mechanisms are known as instantons~\cite{Schafer:1995pz,Diakonov:2002fq} and sphalerons~\cite{Arnold:1987zg,Fukugita:1990gb}, respectively.  As temperature ($T$) increases, the sphaleron contribution prevails over that of the instanton, since the tunneling rate is diminished as $\propto\exp(-T^{2})$ by the screening~\cite{Schafer:1995pz}. However, since the instanton contribution remains still considerable below the chiral restoration $T$, $T^{\chi}_{c}\approx200$ MeV, which can be lowered more with the inclusion of dynamical quarks~\cite{Diakonov:1988my,Maezawa:2007fd,Ali Khan:2000iz}, one may expect a sizable instanton contribution for the nontrivial $Q_{\mathrm{t}}$ at low $T$. 

Now, we want to mention the recent discussions on how to measure this interesting QCD vacuum property, demonstrated by the nontrivial $Q_{\mathrm{t}}$, in experiments. In Refs.~\cite{Voloshin:2004vk}, it was suggested that the nontrivial $Q_{\mathrm{t}}$ can be probed by measuring asymmetric electric currents, due to an event-by-event $P$- and $CP$-violation, in the non-central heavy-ion collision experiments such as RHIC, FAIR, and LHC. Their main idea is that unequal numbers of the chirally left- and right-handed quarks, $N_{\mathrm{L}}-N_{\mathrm{R}}\ne0$ proportional to the $Q_{\mathrm{t}}$ in terms of the axial Ward-Takahashi identity~\cite{Kharzeev:2007jp}, can produce the asymmetric electric current under a strong magnetic field $B$, which is generated by the collision and perpendicular to the reaction plane. Interestingly enough, a possible experimental evidence, which is the electric charge separation, was reported recently by STAR collaboration at RHIC~\cite{Voloshin:2008jx}. Moreover, in Ref.~\cite{Buividovich:2009wi}, using the tadpole-improved SU($2$) quenched lattice simulation, the electromagnetic (EM) current fluctuation as well as local chirality were explored at finite $T$, resulting in that the fluctuations become insensitive to the external magnetic field as $T$ increases. Moreover, they supported an instanton-like gauge configuration for the CME at low $T$, since the longitudinal component of the $j_{\mu}$ is drastically enhanced by it. In Ref.~\cite{Fukushima:2008xe}, it was also shown in four different ways that the electric current, induced by the external magnetic field with the nontrivial $Q_{\mathrm{t}}$, reads 
\begin{equation}
\label{eq:EMC}
{\bm j}=-\frac{\mu_{\chi}}{2\pi^{2}}{\bm B},
\end{equation}
where the $\mu_{\chi}$ represents the chiral chemical potential and contains the information on the $Q_{\mathrm{t}}$ inside of it. Note that we set the quark electric charge to be unity in Eq.~(\ref{eq:EMC}) for simplicity. In fact, this is a consequence of the EM axial anomaly~\cite{DHoker:1985yb}.

In the present work, as mentioned above, we investigate the CME at low $T$, employing the instanton vacuum configuration with the nontrivial $Q_{\mathrm{t}}$, which relates to the finite instanton-number fluctuation, {\it i.e.}  the number difference between the instantons and anti-instantons in the grand canonical ensemble, $\Delta\equiv N_{I}-N_{\bar{I}}\ne0$ resulting in the $CP$-violation. To this end, we first write the effective action derived from the instanton vacuum configuration as a functional of $\Delta$~\cite{Diakonov:1995qy}. Using generic functional external-source and linear Schwinger methods, we compute the relevant physical quantities, the vacuum expectation values of the local chiral density $\langle \rho_{\chi}\rangle$, chiral charge density $\langle n_{\chi}\rangle$, EM current $\langle j_{\mu} \rangle$, induced by the external magnetic field, as the signals of the CME. In order to consider the $T$-dependence of the relevant quantities, we also employ the Harrington-Shepard caloron to obtain the $T$-dependent constituent-quark mass. Fianlly, the standard fermionic Matsubara formula is taken into account for the anti-perioidc sum over the Euclidean time, {\it i.e.}  $T$. We consider all the relevant physical quantities up to $\mathcal{O}(\Delta^{2})\sim\mathcal{O}(Q^{2}_{\mathrm{t}})$.

As a result, we can obtain the correct expression for the induced EM current as given in Eq.~(\ref{eq:EMC}) and observe that the longitudinal component of the EM current is much larger than the transverse one, satisfying the relation $|j_{\perp}/j_{\parallel}|\propto Q_{\mathrm{t}}$. Moreover, it turns out that the $\langle n_{\chi}\rangle $ equals to the $|\langle j_{3,4}\rangle|$ as long as the Lorentz invariance remains unharmed. These results are compatible and consistent with those given in Refs.~\cite{Fukushima:2008xe,Buividovich:2009wi}. We also find that the CME becomes insensitive to the magnetic field as $T$ increases, according to the decreasing instanton, {\it i.e.}  tunneling effect, as observed in the lattice QCD simulation~\cite{Buividovich:2009wi}. Within our low-$T$ instanton framework, the instanton contribution to the CME becomes almost negligible beyond $T\approx300$ MeV.

We organize the present work as follows: In Section II, we make a brief discussion on the effective action derived from the instanton vacuum configuration with the finite instanton-number fluctuation $\Delta\ne0$. The relevant physical quantities, which signal the CME, are defined and computed with the effective action in Section III. All the ingredients obtained in the last Sections are extended to the case at finite $T$ in Section IV using the Matsubara formula and the Harrington-Shepard caloron. In Section V, numerical results and related discussions are given. Finally, Section VI is devoted to summary and conclusion. 
\section{Effective action from the instanton vacuum}
In this Section, we briefly introduce a $P$- and $CP$-violating effective action $\mathcal{S}_{\mathrm{eff}}$, derived by Diakonov {\it et al.} from the instanton vacuum configuration in the large $N_{c}$ limit at zero temperature ($T=0$)~\cite{Diakonov:1995qy}. Employing a dilute grand canonical ensemble of the (anti)instantons with a finite instanton-number fluctuation, $\Delta\equiv N_{+}-N_{-}\ne0$, which corresponds to a $CP$-violating vacuum, but a fixed total number of the pseudo-particles $N_{+}+N_{-}=N$, the $\mathcal{S}_{\mathrm{eff}}$ can be written in momentum space with Euclidean metric as follows:
\begin{eqnarray}
\label{eq:EA}
\mathcal{S}_{\mathrm{eff}}
&=&\mathcal{C}+\frac{N_{+}}{V}\ln\lambda_{+}
+\frac{N_{-}}{V}\ln\lambda_{-}
-\frac{mN_{c}}{4\pi^{2}\bar{\rho}^{2}}(\lambda_{+}+\lambda_{-})
\cr
&-&N_{c}\int\frac{d^{4}k}{(2\pi)^{4}}\mathrm{Tr}_{\gamma}\ln
\left[\frac{\rlap{/}{k}-\frac{i}{2}
[\lambda_{+}(1+\gamma_{5})+\lambda_{-}(1-\gamma_{5})]F^{2}(k)}
{\rlap{/}{k}-im}\right],
\end{eqnarray}
where we have used $N_{f}=1$. However, the extension to an arbitrary $N_{f}$ is just straightforward. The $\mathcal{C}$ stands for an irrelevant constant for further investigations, whereas the $N_{\pm}/V$ for the (anti)instanton packing fraction proportional to the inverse of the average inter-(anti)instanton distance $1/\bar{R}^{4}\approx(200\,\mathrm{MeV})^{4}$. The $\lambda_{\pm}$ denotes a Lagrangian multiplier, which was employed to exponentiate the $2N_{f}$-'t Hooft interaction in the effective action~\cite{Diakonov:2002fq}. The average instanton size in the dilute instanton ensemble is assigned as $1/\bar{\rho}\approx600$ MeV, while the $m$ indicates a small but finite current-quark mass for the SU(2) light-flavor sector ($m\to0$). The $F(k)$ denotes the quark form factor originating from the non-local quark-instanton interactions and is defined as
\begin{equation}
\label{eq:FF}
F(k)=\frac{t}{2}{\rho}\left[I_{0}(t)K_{1}(t)-I_{1}(t)K_{0}(t)
-\frac{1}{t}I_{1}(t)K_{1}(t) \right],\,\,\,\,t=\frac{|k|\bar{\rho}}{2},
\end{equation}
where the $I_{n}$ and $K_{n}$ stand for the modified Bessel functions. We, however, will employ a parameterization of this from factor for convenience in the numerical calculations. 

From the effective action, we can obtain the following two self-consistent (saddle-point) equations with respect to the $\lambda_{\pm}$:
\begin{eqnarray}
\label{eq:SDP}
\lambda_{\pm}
\frac{\partial \mathcal{S}_{\mathrm{eff}}}{\partial \lambda_{\pm}}
&=&\frac{N_{\pm}}{V}-\frac{\lambda_{\pm}mN_{c}}{4\pi^{2}\bar{\rho}^{2}}
+N_{c}\int\frac{d^{4}k}{(2\pi)^{4}}\mathrm{Tr}_{\gamma}
\left[\frac{\frac{i\lambda_{\pm}}{2}
(1\pm\gamma_{5})F^{2}(k)}{\rlap{/}{k}-\frac{i}{2}
[\lambda_{+}(1+\gamma_{5})+\lambda_{-}(1-\gamma_{5})]F^{2}(k)}\right]
\cr
&=&\frac{N_{\pm}}{V}-\frac{(1\pm\delta)M_{0}mN_{c}}
{4\pi^{2}\bar{\rho}^{2}}
-N_{c}\int\frac{d^{4}k}{(2\pi)^{4}}\mathrm{Tr}_{\gamma}
\left[\frac{\frac{1}{2}(1\pm\gamma_{5})
(1+\delta\gamma_{5})^{2}M^2}
{k^{2}+(1+\delta\gamma_{5})^{2}M^2}\right]=0,
\end{eqnarray}
where the $\lambda_{\pm}$ is approximated as $M_{0}(1\pm\delta)$ in the last line of Eq.~(\ref{eq:SDP}) with account of the fact $\Delta\ll N$ in the thermodynamic limit~\cite{Diakonov:1995qy}. The momentum-dependent constituent-quark mass is defined as $M(k)=M_{0}F^{2}(k)$~\cite{Diakonov:2002fq}. By adding and subtracting the instanton $(+)$ and anti-instanton ($-$) contributions in Eq.~(\ref{eq:SDP}), we arrive at
\begin{equation}
\label{eq:NOV}
\frac{N}{V}-\frac{mM_{0}N_{c}}
{2\pi^{2}\bar{\rho}^{2}}\approx4N_{c}\int\frac{d^{4}k}{(2\pi)^{4}}
\frac{(1+\delta^{2})M^{2}}
{k^{2}+(1+\delta^{2})M^{2}},
\end{equation}
\begin{equation}
\label{eq:DOV}
\frac{\Delta}{V}-\frac{\delta mM_{0}N_{c}}
{2\pi^{2}\bar{\rho}^{2}}\approx8N_{c}\int\frac{d^{4}k}{(2\pi)^{4}}
\frac{\delta M^{2}}
{k^{2}+(1+\delta^{2})M^{2}}.
\end{equation}
Taking into account that $\delta\ll1$, $\Delta\ll N$, and using Eq.~(\ref{eq:DOV}), we can obtain an expression for $\delta$ as a function of relevant parameters:
\begin{equation}
\label{eq:DEL}
\delta=\left(\frac{2\pi^{2}\bar{\rho}^{2}}{mM_{0}N_{c}}\right)
\frac{\Delta}{V}.
\end{equation}
This equation tells us that the $\delta$ contains the information on the instanton-number fluctuation $\Delta$ at a certain scale $\bar{\rho}$, which is about $600$ MeV in the present framework. Taking into account all the ingredients discussed so far, finally, we can write the relevant effective action with $\Delta\ne0$ for further investigations:
\begin{eqnarray}
\label{eq:EA2}
\mathcal{S}_{\mathrm{eff}}
&=&-\int\frac{d^4k}{(2\pi)^4}\mathrm{Tr}_{c,f,\gamma}\ln
\left[\frac{\rlap{/}{k}-i(1+\delta\gamma_{5})M(k)}
{\rlap{/}{k}-im}\right],
\end{eqnarray}
where the $\mathrm{Tr}_{c,f,\gamma}$ denotes the trace over color, flavor  and Lorentz indices. 

Now, we are in a position to discuss the relation between the nontrivial topological charge $Q_{\mathrm{t}}$, as a source of the CME, and the instanton number fluctuation $\Delta$. According to the axial Ward-Takahashi identity~\cite{Kharzeev:2007jp}, the $Q_{\mathrm{t}}$ is proportional to the number difference between the chirally left- and right-handed quarks, $Q_{\mathrm{t}}\propto N_{R}-N_{L}$. Hence, the nontrivial $Q_{\mathrm{t}}$ indicates the chirality flip. Note that, similarly, if a chirally left-handed quark is scattered from an instanton to an anti-instanton, the quark helicity is flipped to the right-handed one, and vice versa. This means that the nonzero $\Delta$ results in $N_{R}-N_{L}\ne0$. In this way, the $Q_{\mathrm{t}}$ can be considered to be proportional to the $\Delta$: $Q_{\mathrm{t}}\sim\Delta$~\cite{Schafer:1996wv,Diakonov:2002fq}. As a consequence, we can study the CME using the effective action in Eq.~(\ref{eq:EA2}) as a functional of $Q_{\mathrm{t}}$, more explicitly $\delta\propto\Delta$. The $\delta$ in Eq.~(\ref{eq:DEL}) can be rewritten in terms of a real and small parameter $\epsilon$, which satisfies the condition $|\epsilon|\leq1$, for convenience as follows:
\begin{equation}
\label{eq:44}
\delta=\left(\frac{2\pi^{2}\bar{\rho}^{2}}{mM_{0}N_{c}}\right)
\frac{\epsilon N}{V},
\end{equation}
Using the typical values for the parameters for the flavor SU(2) instanton model listed in Table~\ref{TABLE0}, one can obtain $\delta\approx13.57\epsilon$. 

\begin{table}[h]
\begin{tabular}{c|c|c|c|c}
$\bar{R}$
&$\bar{\rho}$
&$(N/V)^{1/4}$
&$M_{0}$
&$m$\\
\hline
$1$ fm&$1/3$ fm&$200$ MeV
&$350$ MeV&$5$ MeV\\
\end{tabular}
\caption{Instanton parameters, constituent and current quark masses in vacuum.}
\label{TABLE0}
\end{table}
\section{The Chiral magnetic effect in the magnetic field}
In this Section, we compute relevant physical quantities for the CME in the presence of the external magnetic field, using the effective action given in the last Section. They are the vacuum expectation values (v.e.v.) of the local chiral density $\rho_{\chi}$, chiral charge density $n_{\chi}$, and electromagnetic (EM) current $j_{\mu}$, which are defined as follows:
\begin{equation}
\label{eq:DEF}
\rho_{\chi}(x)=iq^{\dagger}(x)\gamma_{5}q(x),\,\,\,\,
n_{\chi}(x)=iq^{\dagger}(x)\gamma_{4}\gamma_{5}q(x),\,\,\,\,
j_{\mu}(x)=iq^{\dagger}(x)\gamma_{\mu}q(x).
\end{equation}
The $\rho_{\chi}$ and $n_{\chi}$ represent the strength of the parity breaking in a system, whereas the $j_{\mu}$ here the EM current induced by the external magnetic field. It is very convenient to compute these quantities employing the effective action with corresponding pseudoscalar ($\mathcal{P}$), axial vector ($\mathcal{A}$), and  vector ($\mathcal{V}$) external sources. Hence, the effective action can be rewritten as: 
\begin{eqnarray}
\label{eq:EAV}
\mathcal{S}_{\mathrm{eff}}
&=&-\int\frac{d^4k}{(2\pi)^4}\mathrm{Tr}_{c,f,\gamma}\ln
\left[\frac{\rlap{/}{K}-i(1+\delta\gamma_{5})M(K)
+i\delta\gamma_{5}\mathcal{P}
+\delta\gamma_{4}\gamma_{5}\mathcal{A}_{4}
+\gamma_{\mu}\mathcal{V}_{\mu}}
{\rlap{/}{K}-im+i\delta\gamma_{5}\mathcal{P}
+\delta\gamma_{4}\gamma_{5}\mathcal{A}_{4}
+\gamma_{\mu}\mathcal{V}_{\mu}}\right].
\end{eqnarray}
We will take $N_{c}=3$ and $N_{f}=2$ throughout the present work. The $K_{\mu}$ indicates the covariant quark momentum, gauged by the photon field as $k_{\mu}+A_{\mu}$, in which we set the quark electric charge unity for convenience. Note that we have written the external sources $\mathcal{P}$ and $\mathcal{A}_{4}$, which correspond to the $\rho_{\chi}$ and $n_{\chi}$, to be proportional to the $\delta$, since they signal the nontrivial topological charge $Q_{\mathrm{t}}$ as the parity-breaking quark mass $\delta\gamma_{5}M$ does.  
 
First, we calculate the v.e.v. of the chiral density, using the standard functional method as follows: 
\begin{equation}
\label{eq:CD1}
\langle\rho_{\chi}\rangle
=\frac{\partial\mathcal{S}_{\mathrm{eff}}}
{\partial\mathcal{P}}
=-i\delta N_{c}N_{f}\int\frac{d^{4}k}{(2\pi)^{4}}\mathrm{Tr}_{\gamma}
\left\{\left[
\frac{1}{\rlap{/}{K}-i(1+\delta\gamma_{5})M(K)}
-\frac{1}{\rlap{/}{K}-im}\right]\gamma_{5} \right\},
\end{equation}
where we have performed the trace of over the color and flavor indices. Since we are interested in the response of the nonperturbative vacuum to  the external magnetic field for the nonzero topological charge $Q_{\mathrm{t}}$, we only collect the terms proportional to the $\delta$ and the photon field strength tensor $F_{\mu\nu}$. In order to perform the trace over the Lorentz index in the r.h.s. of Eq.~(\ref{eq:CD1}) under the external EM field, we employ the linear Schwinger method~\cite{Schwinger:1951nm,Nieves:2006xp}. According to that, the quark propagator in the presence of the instanton background can be written up to $\mathcal{O}(F_{\mu\nu})$, which is equivalent up to $\mathcal{O}(B_{0})$, as follows:
\begin{eqnarray}
\label{eq:DEEX}
S(k,A)&=&\frac{1}{\rlap{/}{K}-i(1+\delta\gamma_{5})M(K)}
\approx
\frac{\rlap{/}{k}+\rlap{/}{A}
+i(1+\delta\gamma_{5})\left[M+\frac{1}{2}\bar{M}
(\sigma\cdot F)\right]}
{k^{2}+(1+\delta^{2})M^{2}}
\cr
&\times&
\left[1-\frac{\tilde{M}(\sigma\cdot F)+i\left(1-\frac{\delta}{4}\gamma_{5} \right)\hat{M}(k)\gamma_{\mu}K_{\nu}F_{\mu\nu}-2i\delta M\gamma_{5}\rlap{/}{K}}{k^{2}+(1+\delta^{2})M^{2}}\right].
\end{eqnarray}
Here, $\sigma\cdot F=\sigma_{\mu\nu}F_{\mu\nu}$. We define relevant functions related to the momentum-dependent quark mass:
\begin{equation}
\label{eq:MF}
M=M_{0}\left(\frac{2}{2+k^{2}\bar{\rho}^{2}} \right)^{2},\,\,\,\,
\bar{M}=-\frac{8M_{0}\bar{\rho}^{2}}{(2+k^{2}\bar{\rho}^{2})^{3}},\,\,\,\,
\tilde{M}=\frac{1}{2}+M\bar{M},\,\,\,\,\hat{M}=4i\bar{M}.
\end{equation}
Note that these functions are appropriate parameterizations from Eq.~(\ref{eq:FF}) as mentioned in the last Section. After a straightforward trace manipulation, one arrives at a simple expression for the chiral density:
\begin{equation}
\label{eq:CD2}
\langle\rho_{\chi} \rangle_{F,\delta}
=2\delta^{2}\mathcal{F}_{a}B^{2}_{0},
\end{equation}
where the subscripts $F$ and $\delta$ in the l.h.s. of Eq.~(\ref{eq:CD2}) indicate the fact that we picked up only the terms proportional to the $F_{\mu\nu}$ and $\delta$ as explained above. In deriving Eq.~(\ref{eq:CD2}), to make the problem easy, we assumed a static external EM field, which is assigned for example as in Ref.~\cite{Buividovich:2009wi}: 
\begin{equation}
\label{eq:A}
A_{\mu}=A^{\mathrm{cl}}_{\mu}+A^{\mathrm{fluc}}_{\mu}
=\left(-\frac{B_{0}}{2}x_{2},\frac{B_{0}}{2}x_{1},0,0 \right)+(a1,a2,a3,a4),
\end{equation}
where the external EM field consists of the classical ($A^{\mathrm{cl}}_{\mu}$) in the symmetry gauge and fluctuation ($A^{\mathrm{fluc}}_{\mu}$) parts. We take the $a_{1\sim4}$ as small and constant EM potentials, satisfying $\partial_{\mu}a_{1\sim4}=0$. In this assignment for the $A_{\mu}$, we have only the constant magnetic field in the spatial $\hat{x}_{3}$-direction: ${\bm B}=B_{0}\hat{x}_{3}$. Thus, among the EM field strength tensors, $F_{12}=B_{0}$ and its dual remain finite, while others disappear. Note that we define a dual field strength tensor in Euclidean space as $\tilde{F}_{\mu\nu}=\frac{1}{2}\epsilon_{\mu\nu\rho\sigma}F_{\rho\sigma}$. The relevant integral $\mathcal{F}_{a}$ in Eq.~(\ref{eq:CD2}) reads:
\begin{equation}
\label{eq:Fa}
\mathcal{F}_{a}=-4N_{c}N_{f}\int\frac{d^{4}k}{(2\pi)^{4}}
\frac{\bar{M}\left(M\bar{M}+\frac{1}{2} \right)}
{[k^{2}+(1+\delta^{2})M^{2}]^{2}}, 
\end{equation}
As understood from Eq.~(\ref{eq:CD2}), the parity breaking of the vacuum becomes enhanced quadratically with respect to the $\delta$ and $B_{0}$.

Similarly to the $\rho_{\chi}$, we can compute the v.e.v. of the chiral charge density using the following functional derivative with respect to the external axial-vector field:
\begin{equation}
\label{eq:CCD1}
\langle{n_{\chi}}\rangle=
\frac{\partial\mathcal{S}_{\mathrm{eff}}}
{\partial\mathcal{A}_{4}}=
-\delta N_{c}N_{f}\int\frac{d^{4}k}{(2\pi)^{4}}\mathrm{Tr}_{\gamma}
\left\{\left[
\frac{1}{\rlap{/}{K}-i(1+\delta\gamma_{5})M(K)}
-\frac{1}{\rlap{/}{K}-im}
\right]\gamma_{4}\gamma_{5}\right\}.
\end{equation}
By solving Eq.~(\ref{eq:CCD1}) and picking up relevant terms, one is lead to a compact expression for it as follows:
\begin{equation}
\label{eq:CCD2}
\langle{n_{\chi}}\rangle_{F,\delta}
=\delta\mathcal{F}_{b}\epsilon_{4\nu\rho\sigma}(iA_{\sigma})F_{\nu\rho}
=2\delta\mathcal{F}_{b}(iA_{3})B_{0},
\end{equation}
where we have used the fact that only the $\tilde{F}_{34}=B_{0}$ is nonzero in our symmetric-gauge EM field. From Eq.~(\ref{eq:CCD2}), we observe that the chiral charge density increases linearly with respect to the $\delta$ as well as the $B_{0}$. The integral $\mathcal{F}_{b}$ is written as: 
\begin{equation}
\label{eq:Fb}
\mathcal{F}_{b}=-4N_{c}N_{f}\int\frac{d^4k}{(2\pi)^4}
\frac{M\bar{M}}
{[k^{2}+(1+\delta^{2})M^{2}]^{2}}.
\end{equation}
Finally, we attempt to compute the v.e.v. of the EM current, induced by the external magnetic field in the presence of the nontrivial topological charge:
\begin{equation}
\label{eq:EMC1}
\langle{j_{\mu}}\rangle=\frac{\partial\mathcal{S}_{\mathrm{eff}}}
{\partial\mathcal{V}_{\mu}}=-N_{c}N_{f}\int\frac{d^{4}k}{(2\pi)^{4}}
\mathrm{Tr}_{\gamma}
\left\{\left[
\frac{1}{\rlap{/}{K}-i(1+\delta\gamma_{5})M(K)}
-\frac{1}{\rlap{/}{K}-im}
\right]\gamma_{\mu}\right\}.
\end{equation}
Expectedly, we can obtain a very similar expression for the $\langle j_{\mu}\rangle_{F,\delta}$ to that for the v.e.v. of the chiral charge density:
\begin{eqnarray}
\label{eq:EMC2}
\langle{j_{\mu}}\rangle_{F,\delta}=
-[3\delta^{2}F_{\mu\nu}+2\delta\tilde{F}_{\mu\nu}](iA_{\nu})\mathcal{F}_{b},
\end{eqnarray}
where the definition of the $\mathcal{F}_{b}$ is the same with that given in Eq.~(\ref{eq:Fb}). Considering Eq.~(\ref{eq:EMC2}) and assuming finite values for $A_{1\sim4}$, we can write the induced EM currents for $Q_{\mathrm{t}}\ne0$ separately for its transverse ($\perp$) and longitudinal ($\parallel$) components: 
\begin{eqnarray}
\label{eq:JOA}
\langle{j_{1,2}}\rangle_{F,\delta}
&\equiv&
j_{\perp}=\mp3\delta^{2}\mathcal{F}_{b}(iA_{2,1})B_{0},
\cr
\langle{j_{3,4}}\rangle_{F,\delta}
&\equiv&
j_{\parallel}=\mp2\delta\mathcal{F}_{b}(iA_{4,3})B_{0}.
\end{eqnarray}
The $j_{\parallel}$ increases linearly as the $B_{0}$ grows, which is the indication of the CME as in Eq.~(\ref{eq:EMC}). Equating the transverse and longitudinal components, we have the ratio as follows:
\begin{equation}
\label{eq:JJJ}
\left|\frac{j_{\perp}}{j_{\parallel}}\right|=\frac{3}{2}\frac{|A_{2,1}|}{|A_{4,3}|}\delta.
\end{equation}
From Eq.~(\ref{eq:JJJ}), if we assume that $|A_{1,2}|\sim|A_{3,4}|$ along the $\hat{x}_{3}$-direction ($x_{1}\sim x_{2}\sim0$), in other words, the strengths of the fluctuations $a_{1\sim4}$ in Eq.~(\ref{eq:A}) are almost the same to each other, it is obvious that the transverse components of the induced EM current are much smaller than those of the longitudinal ones, $|j_{\perp}/j_{\parallel}|\ll1$, due to the fact that $\delta\ll1$. This tendency is just consistent with that given in the recent lattice QCD simulation~\cite{Buividovich:2009wi}, in which the instanton-like contribution turned out to be crucial for this  large difference between the longitudinal and transverse components of the EM current for the nontrivial $Q_{\mathrm{t}}$.

Considering the general expression for the induced EM current in Eq.~(\ref{eq:EMC}), the induced EM current along the $\hat{x}_{3}$-direction in Eq.~(\ref{eq:JOA}) can be rewritten as
\begin{equation}
\label{eq:JJJ1}
\langle j_{3}\rangle_{F,\delta}=
-\frac{1}{2\pi^{2}}[4\pi^{2}\delta\mathcal{F}_{b}(iA_{4})]B_{0},
\end{equation}
and gives the following expression for the chiral chemical potential,
\begin{equation}
\label{eq:JJJ2}
\mu_{\chi}
=4\pi^{2}\mathcal{F}_{b}(i\delta A_{4}).
\end{equation}
Substituting $\bar{\rho}\approx1/3$ fm and $M_{0}\approx0.35$ GeV to the $\mathcal{F}_{b}$ for $\delta\approx0$ in Eq.~(\ref{eq:Fb}), Eq.~(\ref{eq:JJJ2}) can be estimated as $\mu_{\chi}\approx0.92\times(i\delta A_{4})$. As can be seen from the effective action in Eq.~(\ref{eq:EAV}), the term $\delta A_{4}$ plays the role of a chemical potential. Hence, this estimation shows a correct behavior of the $\delta A_{4}$: $\mu_{\chi}=i\delta A_{4}$. About $8\%$ shortage may be understood by the fact that there can be additional contributions to Eq.~(\ref{eq:JJJ2}), if we expand  the propagator in Eq.~(\ref{eq:DEEX}) more to higher orders in $\delta$. Moreover, comparing Eqs.~(\ref{eq:CCD2}) and (\ref{eq:JOA}), it can be easily shown that the v.e.v. of the chiral charge density is the same with the third and fourth components of the induced EM current in the leading contributions in $\delta$:
\begin{equation}
\label{eq:NA}
\langle n_{\chi} \rangle_{F,\delta}=\mp\langle j_{3,4} \rangle_{F,\delta}
\approx\frac{1}{2\pi^{2}}(i\delta A_{4,3})B_{0}
=\frac{1}{2\pi^{2}}\mu_{\chi}B_{0},
\end{equation}
where we have used the result from Eq.~(\ref{eq:JJJ2}). Eq.~(\ref{eq:NA}) is consistent with that given in Ref.~\cite{Sadooghi:2006sx,Fukushima:2008xe} for a homogeneous magnetic field, if $A_{3}=A_{4}$. 

So far, we have derived the expressions for three relevant quantities using the effective action. Here is one caveat. In this instanton approach, the interaction of the quarks are nonlocal, resulting in that the vector and axial-vector currents are not conserved in a usual manner~\cite{Nam:2006sx}. To overcome this problem, one has to gauge the quark momentum even in the momentum-dependent quark mass $M$ in Eq.~(\ref{eq:EAV}) with the external vector and axial-vector currents. This treatment will provide additional terms to Eqs.~(\ref{eq:CD2}), (\ref{eq:CCD2}) and (\ref{eq:JOA}). We verified, however, that these terms are odd in the momentum integral, such as $\int d^{4}k\,k_{\mu}f(k^{2})$, if we take only the leading contributions up to $\mathcal{O}(\delta^{2})$. Thus, their contributions can be ignored rather safely, although it is slightly problematic in the current conservation issue for all orders in $\delta$.  

Finally, we close this Section with remarks on the Landau levels and corresponding dimensional reduction, which have not been taken into account in the present work. In the presence of the strong magnetic field $B_{0}\hat{x}_{3}$, the transverse momenta of fermions are quantized so that they are decoupled from the longitudinal components. Especially, as for the lowest Landau level (LLL) approximation, the transverse fermion propagator, which is factorized from that of the longitudinal one, can be written in a Gaussian function, $S^{\mathrm{LLL}}_{\perp}=\exp(-k^{2}_{\perp}/B_{0})$. This makes the calculation of matrix elements greatly simple in such way that the integral over the transverse momenta becomes just $B_{0}/n$, where the $n$ indicates the number of the fermion propagators in a matrix element~\cite{Hong:1996pv,Sadooghi:2006sx}. Hence, one is left with only the integrals over the longitudinal momenta. This is called the dimensional reduction in the strong magnetic field. However, our approach given in this Section did not show the dimensional reduction, since we limited ourselves only up to $\mathcal{O}(\delta^{2})$ in expanding the quark propagator and did not consider the transverse momenta quantization. The present approach may correspond to the effective action method given in Refs.~\cite{DHoker:1985yb,Nieves:2006xp,Fukushima:2008xe}.

\section{Chiral magnetic effect at low temperature}
To investigate the physical quantities in hand at low but finite temperature ($T\lesssim T^{\chi}_{c}$), we want to discuss briefly how to modify the instanton variables, $\bar{\rho}$ and $\bar{R}$ at finite $T$. We will follow our previous work~\cite{Nam:2009nn} and Refs.~\cite{Harrington:1976dj,Diakonov:1988my} to this end. Usually, there are two different instanton configurations at finite $T
$, being periodic in Euclidean time, with trivial and nontrivial holnomies. They are called the Harrington-Shepard (HS)~\cite{Harrington:1976dj} and Kraan-Baal (KB) calorons~\cite{Kraan:1998pm}, respectively. In fact, the nontrivial holonomy can be identified as the Polyakov line as an order parameter for deconfinement phase of QCD. However, since we are not interested in the deconfinement phase in this work, we choose the HS caloron for the modification at finite $T$. Note that here are several caveats; 1) these modifications are done for a pure glue system without dynamical quarks. Hence, the instanton variables may change, if one takes into account the dynamical-quark contributions in the instanton distribution function. 2) Moreover, we assume the $CP$-invariant vacuum for the modifications, {\it i.e.}  $\Delta=0$, whereas we are interested in the physical quantities for $\Delta\ne0$ for the CME. Correcting these inconsistencies may give rise to changes in the final results, although they seem small considering that the order of the $CP$-violation in reality is very tiny, but it must be beyond our scope in the present work, and we want to leave them for future studies. Keeping this issue in mind, we write the instanton distribution function at finite $T$ with the HS caloron as follows:
\begin{equation}
\label{eq:IND}
d(\rho,T)=\underbrace{C_{N_c}\,\Lambda^b_{\mathrm{RS}}\,
\hat{\beta}^{N_c}}_\mathcal{C}\,\rho^{b-5}
\exp\left[-(A_{N_c}T^2
+\bar{\beta}\gamma n\bar{\rho}^2)\rho^2 \right].
\end{equation}
Here, the abbreviated notations are also given as:
\begin{equation}
\label{eq:para}
\hat{\beta}=-b\ln[\Lambda_\mathrm{RS}\rho_\mathrm{cut}],\,\,\,\,
\bar{\beta}=-b\ln[\Lambda_\mathrm{RS}\langle R\rangle],\,\,\,
C_{N_c}=\frac{4.60\,e^{-1.68\alpha_{\mathrm{RS}} Nc}}{\pi^2(N_c-2)!(N_c-1)!},
\end{equation}
\begin{equation}
\label{eq:AA}
A_{N_c}=\frac{1}{3}\left[\frac{11}{6}N_c-1\right]\pi^2,\,\,\,\,
\gamma=\frac{27}{4}\left[\frac{N_c}{N^2_c-1}\right]\pi^2,\,\,\,\,
b=\frac{11N_c-2N_f}{3},\,\,\,\,n=\frac{N}{V}.
\end{equation}
Note that we defined the one-loop inverse charge $\hat{\beta}$ and $\bar{\beta}$ at a certain phenomenological cutoff value $\rho_\mathrm{cut}$ and $\langle R\rangle\approx\bar{R}$. As will be shown, only $\bar{\beta}$ is relevant in the following discussions and will be fixed self-consistently within the present framework. The $\Lambda_{\mathrm{RS}}$ stands for a scale, depending on a renormalization scheme, whereas the $V_3$ for the three-dimensional volume. Using the instanton distribution function in Eq.~(\ref{eq:IND}), we can compute the average value of the instanton size, $\bar{\rho}^2$ straightforwardly as follows~\cite{Schafer:1996wv}:
\begin{equation}
\label{eq:rho}
\bar{\rho}^2(T)
=\frac{\int d\rho\,\rho^2 d(\rho,T)}{\int d\rho\,d(\rho,T)}
=\frac{\left[A^2_{N_c}T^4
+4\nu\bar{\beta}\gamma n \right]^{\frac{1}{2}}
-A_{N_c}T^2}{2\bar{\beta}\gamma n},
\end{equation}
where $\nu=(b-4)/2$. Substituting Eq.~(\ref{eq:rho}) into Eq.~(\ref{eq:IND}), the distribution function can be evaluated further as:
\begin{equation}
\label{eq:dT}
d(\rho,T)=\mathcal{C}\,\rho^{b-5}
\exp\left[-\mathcal{M}(T)\rho^2 \right],\,\,\,\,
\mathcal{M}(T)=\frac{1}{2}A_{N_c}T^2+\left[\frac{1}{4}A^2_{N_c}T^4
+\nu\bar{\beta}\gamma n \right]^{\frac{1}{2}}.
\end{equation}
The instanton number density $n$ can be computed self-consistently as a function of $T$, using the following equation:
\begin{equation}
\label{eq:NOVV}
n^\frac{1}{\nu}\mathcal{M}(T)=\left[\mathcal{C}\,\Gamma(\nu) \right]^\frac{1}{\nu},
\end{equation}
where we have replaced $NT/V_3\to n$, and $\Gamma(\nu)$ indicates the $\Gamma$-fucntion with an argument $\nu$. Note that the $\mathcal{C}$ and $\bar{\beta}$ can be determined easily using Eqs.~(\ref{eq:rho}) and (\ref{eq:NOVV}), incorporating the vacuum values of the $n$ and $\bar{\rho}$: $\mathcal{C}\approx9.81\times10^{-4}$ and $\bar{\beta}\approx9.19$. At the same time, using these results, we can obtain the average instanton size $\bar{\rho}$ as a function of $T$ with Eq.~(\ref{eq:rho}).

\begin{figure}[t]
\begin{tabular}{cc}
\includegraphics[width=7cm]{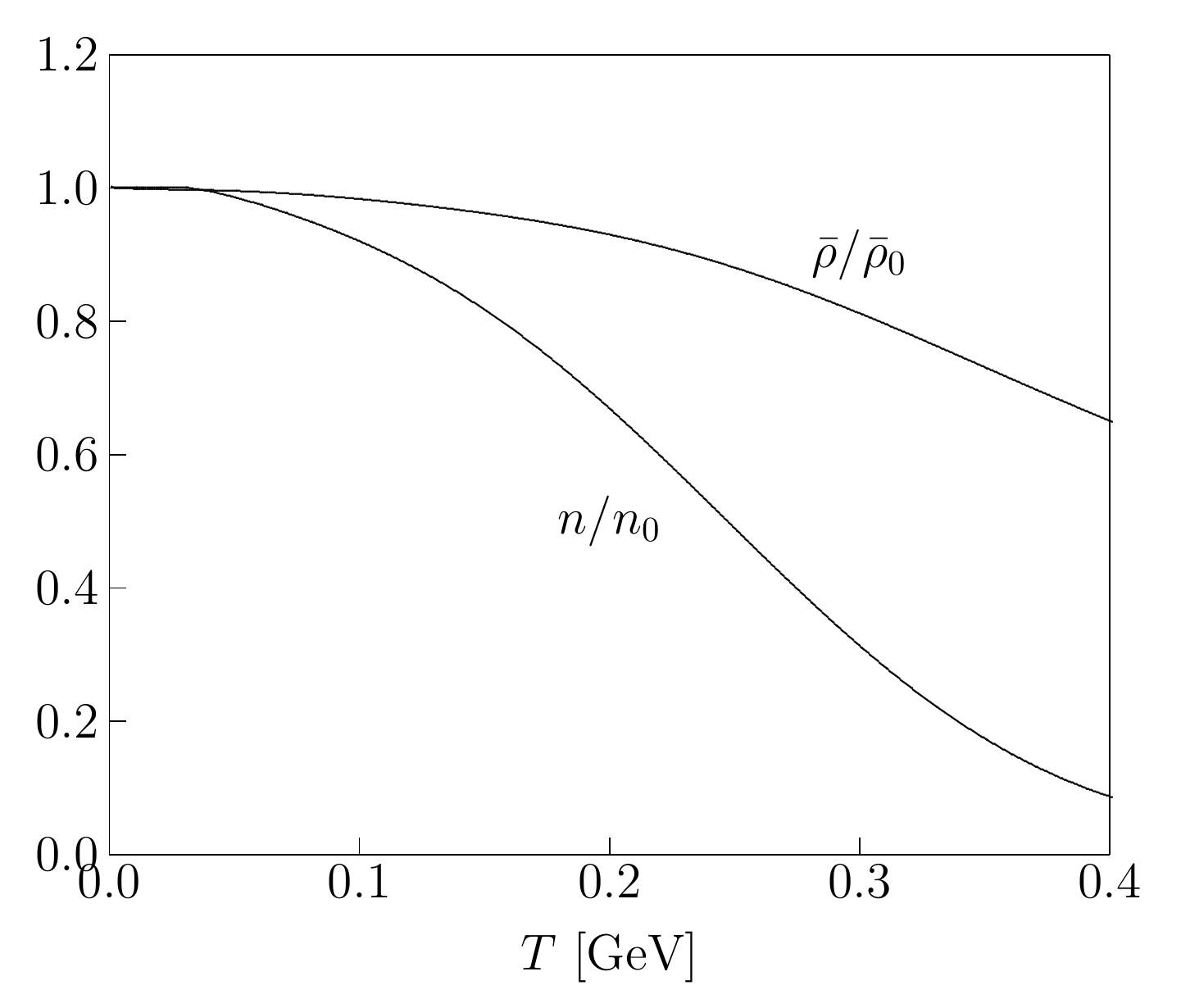}
\includegraphics[width=8cm]{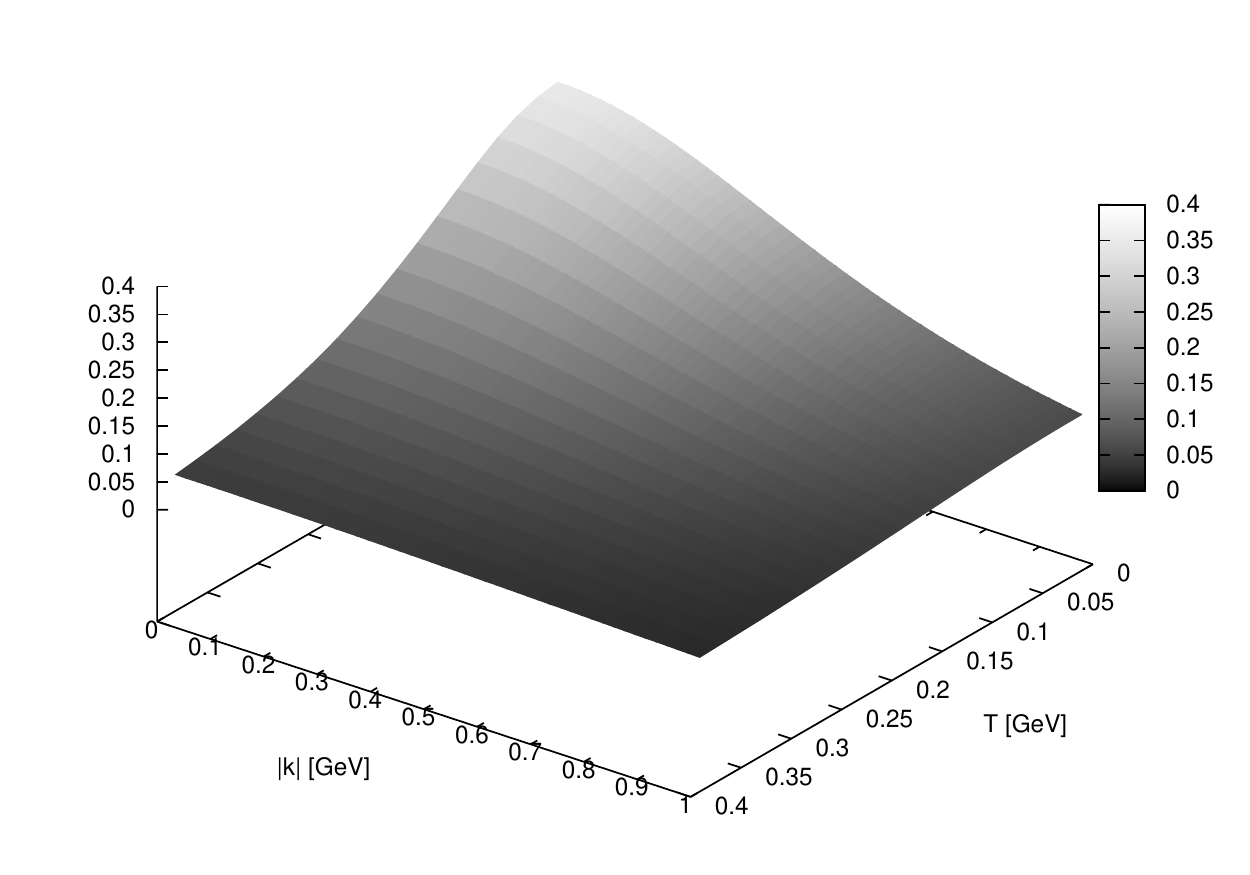}
\end{tabular}
\caption{Left: normalized $\bar{\rho}/\bar{\rho}_{0}$ and $n/n_{0}$ as functions of $T$ for $N_{c}=3$. Right: $M$ as a function of $T$ and absolute value of the momentum $|k|$.}       
\label{FIG1}
\end{figure}

Finally, in order to estimate the $T$-dependence of the constituent-quark mass $M_{0}$, it is necessary to consider the normalized distribution function, defined as follows:
\begin{equation}
\label{eq:NID}
d_N(\rho,T)=\frac{d(\rho,T)}{\int d\rho\,d(\rho,T)}
=\frac{\rho^{b-5}\mathcal{M}^\nu(T)
\exp\left[-\mathcal{M}(T)\rho^2 \right]}{\Gamma(\nu)}.
\end{equation}
Now, we want to employ the large-$N_c$ limit to simplify the expression of $d_N(\rho,T)$. Since the parameter $b$ is in the order of $\mathcal{O}(N_c)$ as shown in Eq.~(\ref{eq:para}), it becomes infinity as $N_c\to\infty$, and the same for $\nu$. In this limit, as understood from Eq.~(\ref{eq:NID}), $d_N(\rho,T)$ can be approximated as a $\delta$-function~\cite{Diakonov:1995qy}: 
\begin{equation}
\label{eq:NID2}
\lim_{N_c\to\infty}d_N(\rho,T)=\delta[{\rho-\bar{\rho}\,(T)}].
\end{equation}
Considering the constituent-quark mass is represented by~\cite{Diakonov:1995qy} 
\begin{equation}
\label{eq:M0}
M_{0}\propto\sqrt{n}\int d\rho\,\rho^{2}\delta[\rho-\bar{\rho}(T)]
=\sqrt{n(T)}\,\bar{\rho}^{2}(T),
\end{equation}
we can modify the $M_{0}$ as a function of $T$ as follows:
\begin{equation}
\label{eq:momo}
M_{0}\to M_{0}\left[\frac{\sqrt{n(T)}\,\bar{\rho}^2(T)}
{\sqrt{n(0)}\,\bar{\rho}^2(0)}\right]\equiv M_{0}(T)
\end{equation}
where we will use $M_{0}\approx350$ MeV as done for zero $T$. The numerical results for the normalized $\bar{\rho}/\bar{\rho}_{0}$ and $n/n_{0}$ as functions of $T$ in the panel of Fig.~\ref{FIG1}. As shown there, these quantities are decreasing with respect to $T$ as expected. However, even beyond the critical $T$ for the chiral restoration $T^{\chi}_{c}\approx\Lambda_{\mathrm{QCD}}\approx200$ MeV, the instanton contribution remains finite. In the right panel of figure, we draw the quark mass as a function of $T$ and absolute value of three momentum of a quark $|\bm{k}|$:
\begin{equation}
\label{eq:M00}
M(|\bm{k}|,T)=M_{0}(T)\left[\frac{2}{2+\bar{\rho}^{2}(T)\,|\bm{k}|^{2}}\right].
\end{equation}
Note that we have ignored the Euclidean-time component of the four momentum in the $M_{\bm{k},T}$ and $\bar{M}_{\bm{k},T}$ by setting $k_{4}=0$. This tricky treatment simplifies the calculations in hand to a large extend, and we also verified that only small deviation appears in comparison to full calculations. Moreover, the $\bar{\rho}$ in Eq.~(\ref{eq:M00}) is now a function of $T$ as demonstrated by Eqs.~(\ref{eq:rho}) and (\ref{eq:momo}) previously. As shown in the figure, the $M(|\bm{k}|,T)$ is a smoothly decreasing function of $T$ and $|\bm{k}|$, indicating that the effect of the instanton is diminished.  For more details, one can refer our previous work~\cite{Nam:2009nn}.

In order to compute the relevant quantities for the CME, given in the previous Section, at finite $T$, it is necessary to compute the $\mathcal{F}_{a}$ and $\mathcal{F}_{b}$ in Eqs.~(\ref{eq:Fa}) and (\ref{eq:Fb}) as functions of $T$. For this purpose, we employ the fermionic Matsubara formula with the following convention:
\begin{equation}
\label{eq:MA}
\int\frac{d^4k}{(2\pi)^4}f(k)\to 
T\sum^{\infty}_{n=-\infty}\int\frac{d^3{\bm k}}{(2\pi)^3}f(w_{n},{\bm k}),
\end{equation}
where the antiperiodic Matsubara frequency is assigned as $w_{n}=(2n+1)\pi T$, where $n\in\mathcal{I}$. Here, we redefine the mass-related functions in Eq.~(\ref{eq:MF}) for the case at finite $T$ as follows:
\begin{equation}
\label{eq:MFT}
M_{\bm{k},T}=M(|\bm{k}|,T),\,\,\,\,
\bar{M}_{\bm{k},T}=-\frac{8M_{0}(T)\,\bar{\rho}^{2}(T)}
{(2+\bar{\rho}^{2}(T)\,|\bm{k}|^{2})^{3}}.
\end{equation}

It is worth mentioning that, in general, the Lorentz invariance is broken down due to the periodic Euclidean time for the direction $\mu=4$, {\it i.e.}  the Matsubara frequency. Hence, the electric part of the field strength ($F_{i4}$ and $F_{4i}$ where $i=1\sim3$) for $T\ne0$ becomes different from that for $T=0$. However, we are free from worrying about such a case in the present work, since we have $F_{i4}=F_{4i}=0$ in our setting of the external EM field as shown in Eq.~(\ref{eq:A}). As a result, the absolute values of the $\langle j_{3}\rangle_{F,\delta}$ and $\langle j_{4}\rangle_{F,\delta}$ in Eq.~(\ref{eq:JOA}) are the same even for $T\ne0$, and the $\langle j_{1}\rangle_{F,\delta}$ and $\langle j_{2}\rangle_{F,\delta}$ also remain unchanged.

Now, we are in a position to compute the $\mathcal{F}_{a,b}$ in Eqs.~(\ref{eq:Fa}) and (\ref{eq:Fb}) as functions of $T$. It is convenient to employ a summation identity for this purpose:
\begin{equation}
\label{eq:iden}
\sum^{\infty}_{n=-\infty}\frac{4T}{(w^{2}_{n}+E^{2}_{\bm{k},T})^{2}}=
\frac{1}{E^{2}_{\bm{k},T}}
\left[\frac{1}{E_{\bm{k},T}}
\frac{(1-e^{-E_{\bm{k},T}/T})}{(1+e^{-E_{\bm{k},T}/T})}
-\frac{1}{2T}
\frac{e^{-E_{\bm{k},T}/T}}{(1+e^{-E_{\bm{k},T}/T})^{2}}
 \right]\equiv\mathcal{M}_{\bm{k},T},
\end{equation}
where we employed a notation $\mathcal{M}_{\bm{k},T}$ for the summation, and the quark energy is written as $E^{2}_{\bm{k},T}={\bm k}^{2}+(1+\delta^{2})M^{2}_{\bm{k},T}$. Then, one arrives finally at
\begin{equation}
\label{eq:Faa}
\mathcal{F}_{a}=
N_{c}\int\frac{d^3{\bm k}}{(2\pi)^3}
\bar{M}_{\bm{k},T}\left(\frac{1}{2}+M_{\bm{k},T}\bar{M}_{\bm{k},T} \right)
\mathcal{M}_{\bm{k},T},
\,\,\,\,
\mathcal{F}_{b}=
N_{c}\int\frac{d^3{\bm k}}{(2\pi)^{3}}
M_{\bm{k},T}\bar{M}_{\bm{k},T}\mathcal{M}_{\bm{k},T}.
\end{equation}
By substituting these $\mathcal{F}_{a,b}$ into Eqs.~(\ref{eq:CD2}), (\ref{eq:CCD2}), and (\ref{eq:EMC2}), we can obtain the v.e.v. of the chiral density, chiral charge density, and the induced EM current as functions of $T$.
\section{Numerical results and Discussions}
\begin{figure}[t]
\begin{tabular}{cc}
\includegraphics[width=7.5cm]{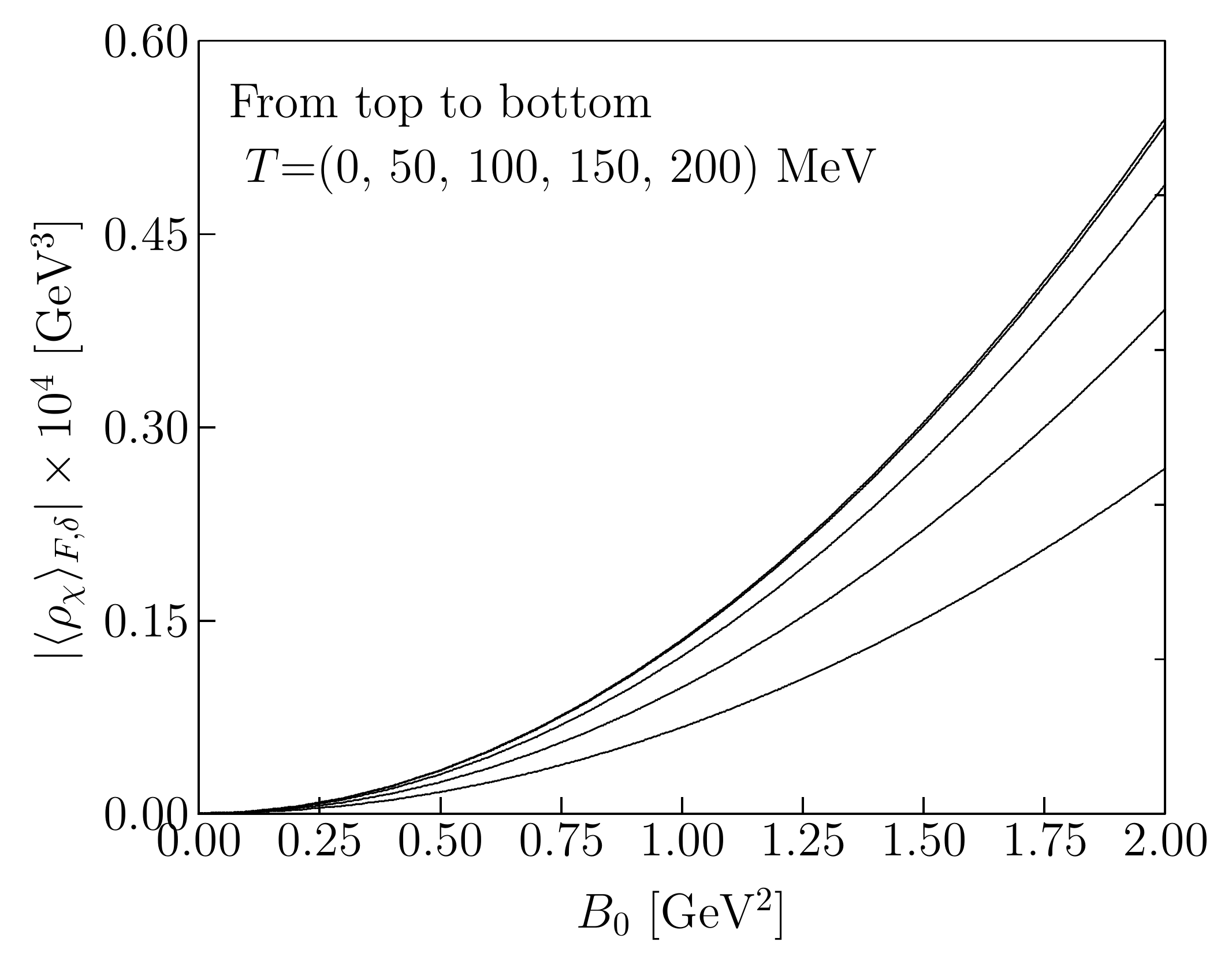}
\includegraphics[width=7.5cm]{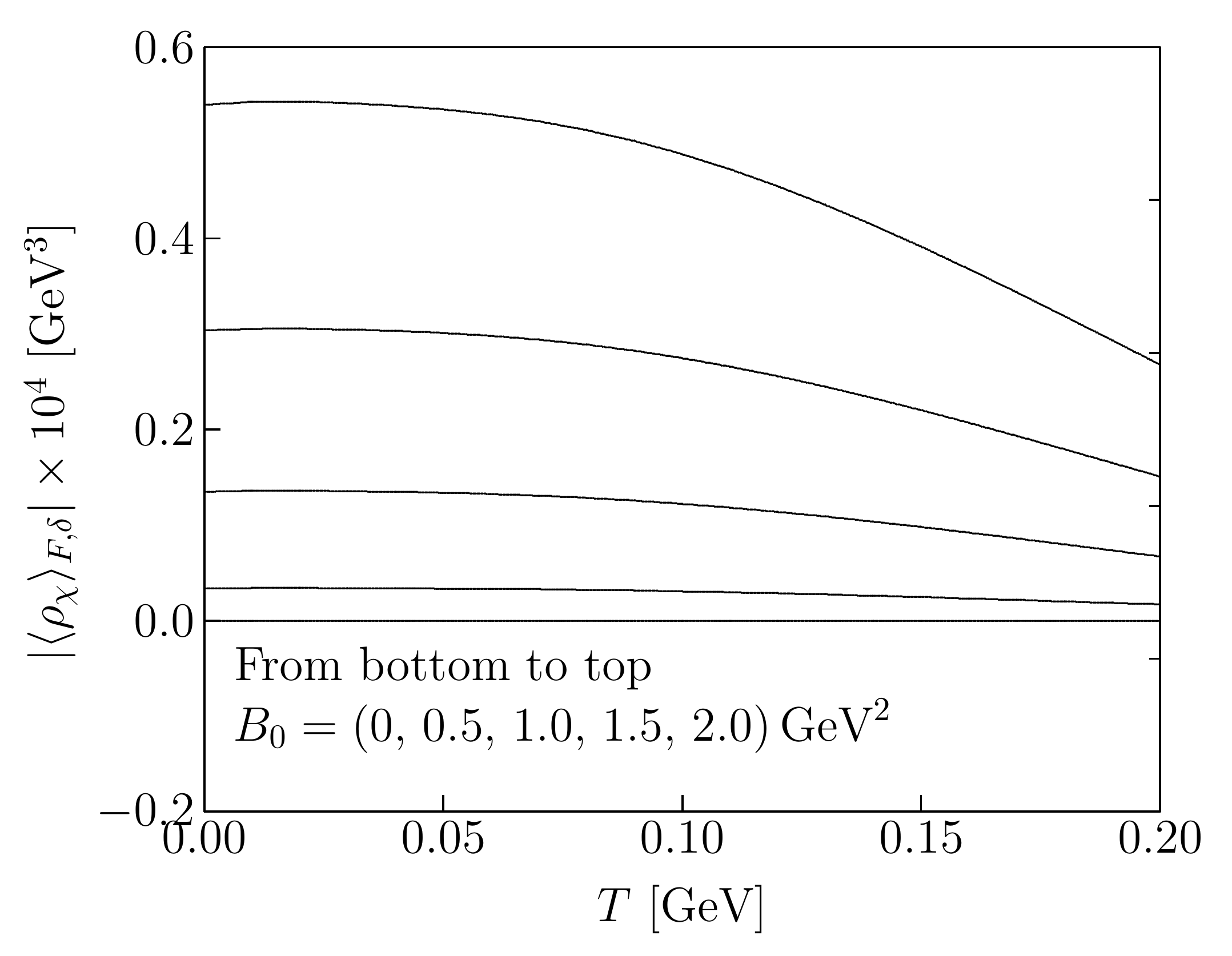}
\end{tabular}
\caption{Chiral local density, $|\langle \rho_{\chi}\rangle_{F,\delta} |\times10^{4}$, as functions of $B_{0}$ (left) and $T$ (right) for $\epsilon=10^{-3}$. The five curves in each panel correspond to those for $T=(0,50,100,150,200)$ MeV (left, from top to bottom) and $B_{0}=(0,0.5,1.0,1.5,2.0)\,\mathrm{GeV}^{2}$ (right, from bottom to top).}       
\label{FIG2}
\end{figure}

In this Section, we present the numerical results for the v.e.v. of the $\langle \rho_{\chi}\rangle$, $\langle n_{\chi}\rangle $, and $\langle j_{\mu}\rangle $ with the nontrivial topological charge $Q_{\mathrm{t}}\ne0$ in the external magnetic field $\bm{B}=B_{0}\hat{x}_{3}$. First, we present the numerical results for the local chiral density $|\langle\rho_{\chi}\rangle_{F,\delta}|$ in Eq.~(\ref{eq:CD2}) as a function of $B_{0}$ in the left panel of Fig.~\ref{FIG2}. There, we depict them for different temperatures, $T=(0,\,50,\,100,\,150,\,200)$ MeV, separately. As a trial, we choose $\epsilon=10^{-3}$, which gives $\delta\approx0.0136$ in Eq.~(\ref{eq:44}). In other words, the $P$- and $CP$-violation effects are about $1\%$ in comparison to non-violating quantities. Moreover, since the $\langle \rho_{\chi}\rangle$ is almost linearly proportional to $\epsilon^{2}$, one can easily estimate it for different $\epsilon$ (or $\delta$) values. Obviously as shown in the figure, the $|\langle \rho_{\chi}\rangle|$ grows rapidly with respect to $B_{0}$, manifesting the CME. As $T$ increases, the strength of the curve decreases. At the same time, the slope of the curve, $\partial|\langle\rho_{\chi}\rangle_{F,\delta}|/\partial B_{0}$, gets smaller. This tendency means that the $|\langle\rho_{\chi}\rangle_{F,\delta}|$ becomes insensitive to the external magnetic field as $T$ gets higher, and the same for the CME. The reason for this tendency can be understood by the decreasing instanton effect.

In the right panel of Fig.~\ref{FIG2}, we show the $|\langle\rho_{\chi}\rangle_{F,\delta}|$ as a function of $T$ for different strengths of the magnetic fields, $B_{0}=(0,\,0.5,\,1.0,\,1.5,\,2.0)\,\mathrm{GeV}^{2}$. As shown there, the strength of a curve depends on that of the magnetic field. As $T$ increases, the $|\langle\rho_{\chi}\rangle_{F,\delta}|$ becomes small, indicating that the instanton effect is reduced. In other words, decreasing tunneling effect corresponding to $Q_{\mathrm{t}}\to0$. Thus, unless there is another mechanism to make the $Q_{\mathrm{t}}$ nontrivial, such as the sphaleron, the $|\langle\rho_{\chi}\rangle_{F,\delta}|$ as well as the CME decrease monotonically with respect to $T$ for a finite $B_{0}$ value. Interestingly, the $|\langle\rho_{\chi}\rangle_{F,\delta}|$ decreases much faster for the larger $B_{0}$ value, since the $T$-dependence of the quantity becomes more obvious and strengthened. 

If we compare the present results to a recent lattice QCD simulation given in Ref.~\cite{Buividovich:2009wi}, in which the correlation of the chiral density $\langle \rho^{2}_{\chi} \rangle$ was taken into account, we observe similarity that the local chiral density becomes insensitive to the magnetic field as $T$ increases: the decreasing slope. However, since we have considered only the contributions proportional to the $B_{0}$ for the $\langle \rho_{\chi} \rangle$, we can not reproduce the nonzero values for it at $B_{0}\approx0$ as shown in Ref.~\cite{Buividovich:2009wi}. We note that, although we did not include in the present numerical calculations, there are additional terms, which are independent on the $B_{0}$ in expanding Eq.~(\ref{eq:CD1}):
\begin{equation}
\label{eq:ADD}
\langle\rho_{\chi}\rangle_{\delta}=
4N_{c}\int\frac{d^4k}{(2\pi)^4}\left[\frac{M}{k^{2}+(1+\delta^{2})M^{2}}
-\frac{8\delta^{2}k^{2}M}{[k^{2}+(1+\delta^{2})M^{2}]^{2}}\right]. 
\end{equation}
Obviously, the first term in the square bracket in the r.h.s. of Eq.~(\ref{eq:ADD}) is just the chiral condensate with the nontrivial $Q_{\mathrm{t}}$.  Note that the second term is far smaller than the first one ($\delta\ll1$). Since the chiral condensate is a smoothly decreasing function of $T$~\cite{Nam:2009nn}, it will give different and nonzero values for the $\langle\rho_{\chi}\rangle_{F,\delta}$ for each $T$ at $B_{0}=0$. 

Now, we are in a position to take into account the $\langle n_{\chi}\rangle$, $\langle j_{\mu}\rangle$, and $\mu_{\chi}$ in Eq.~(\ref{eq:JJJ2}). However, as understood by Eqs.~(\ref{eq:JOA}) and (\ref{eq:NA}), these quantities are related to each other, beside the differences in the constant multiplying-factors in terms of the $\delta$ and $B_{0}$. Hence, it is enough to consider only the $\langle j_{3,4}\rangle_{F,\delta}$ to see the overall behaviors of the quantities. Since the $\langle j_{3,4}\rangle_{F,\delta}$ is functions of undetermined quantities $A_{3,4}$, which have been assumed to be small along the $\hat{x}_{3}$-direction, we compute the absolute value of the normalized quantity $|\langle j_{3,4}\rangle_{F,\delta}/(iA_{3,4})|$ for convenience. 

\begin{figure}[t]
\begin{tabular}{cc}
\includegraphics[width=7.5cm]{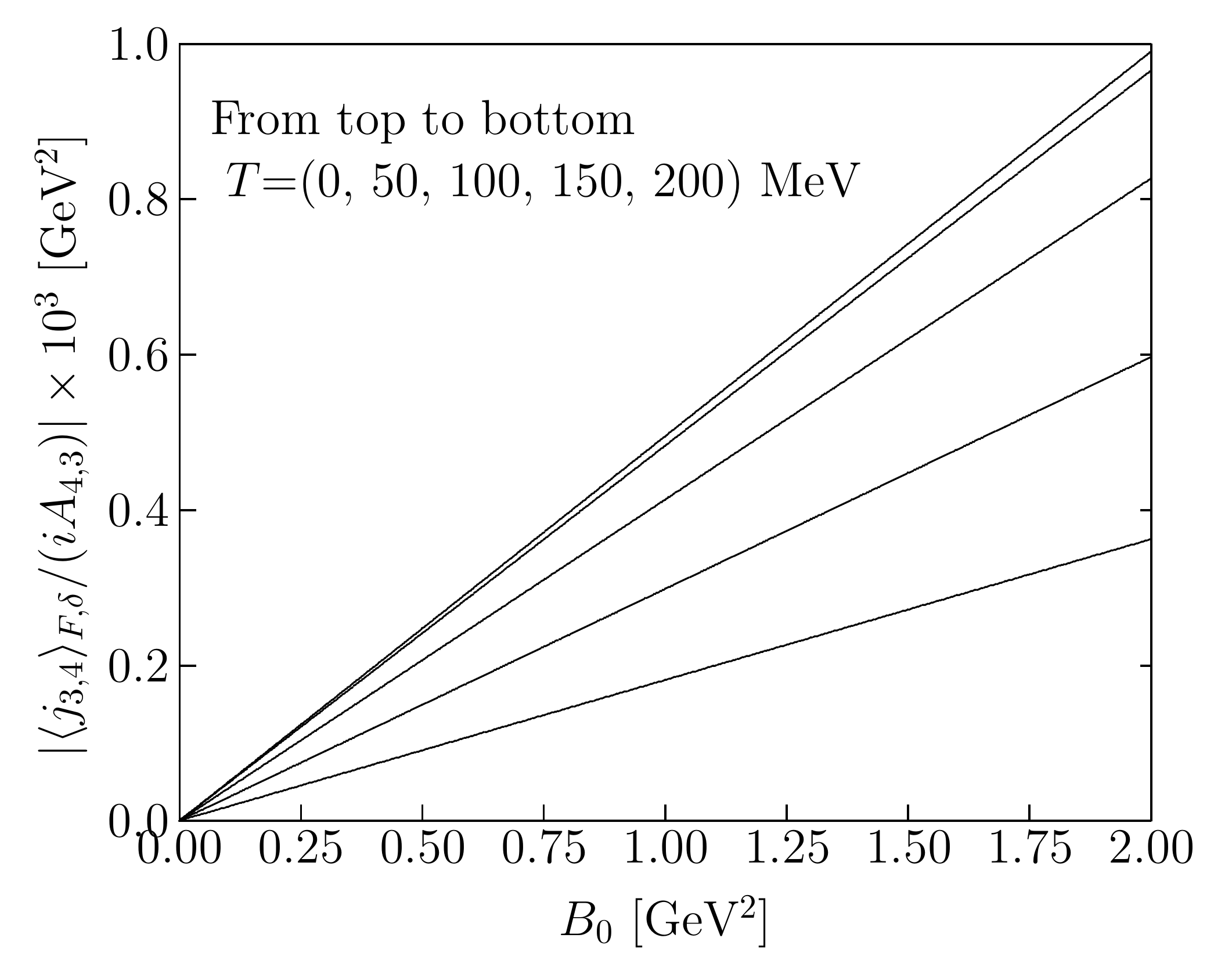}
\includegraphics[width=7.5cm]{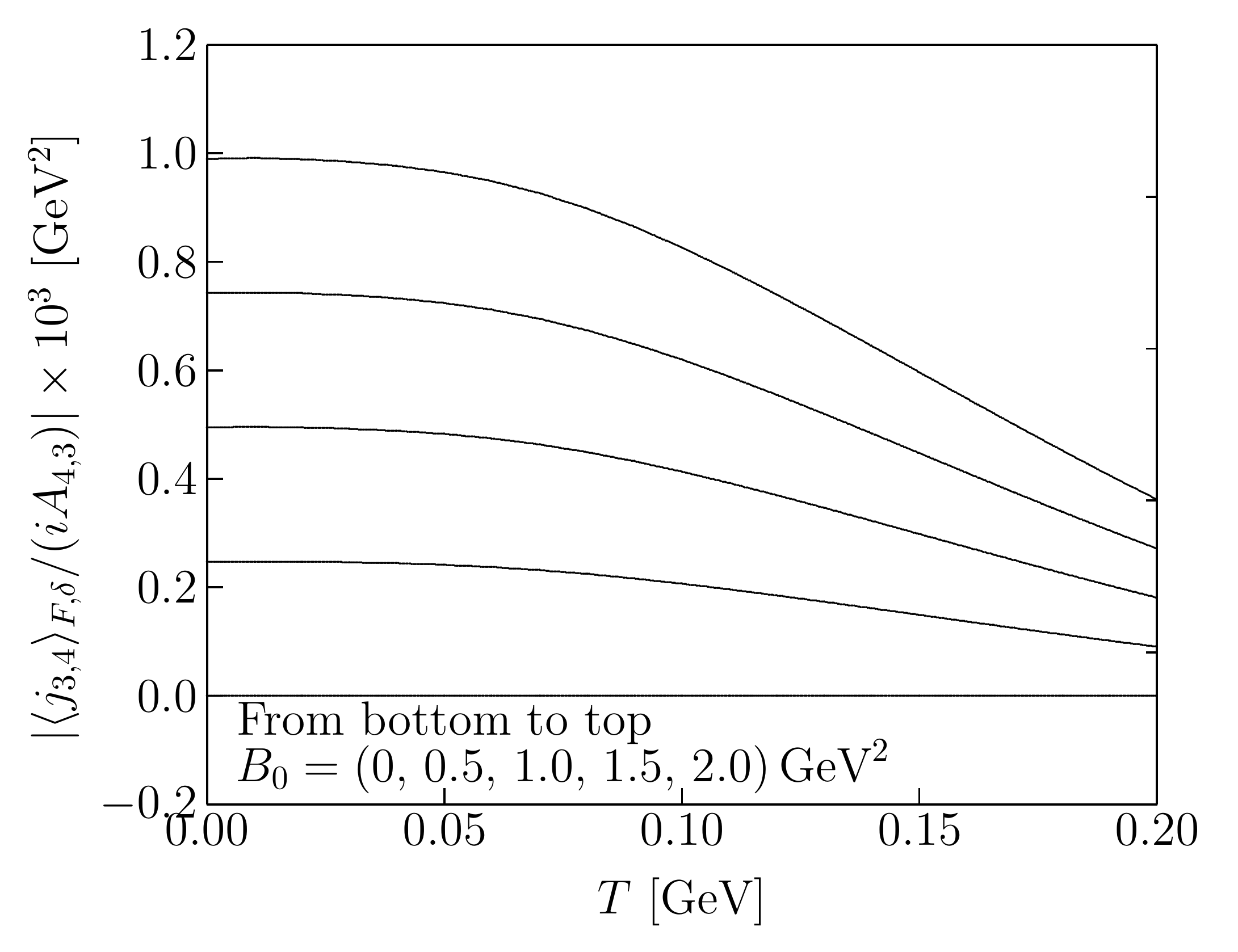}
\end{tabular}
\caption{Induced EM current indicating the CME, $|\langle j_{3,4}/(iA_{4,3})\rangle_{F,\delta}|\times10^{3}$, as functions of $B_{0}$ (left) and $T$ (right) for $\epsilon=10^{-3}$. The five curves in each panel correspond to those for $T=(0,50,100,150,200)$ MeV (left, from top to bottom) and $B_{0}=(0,0.5,1.0,1.5,2.0)\,\mathrm{GeV}^{2}$ (right, from bottom to top).}       
\label{FIG3}
\end{figure}

\begin{figure}[t]
\includegraphics[width=10cm]{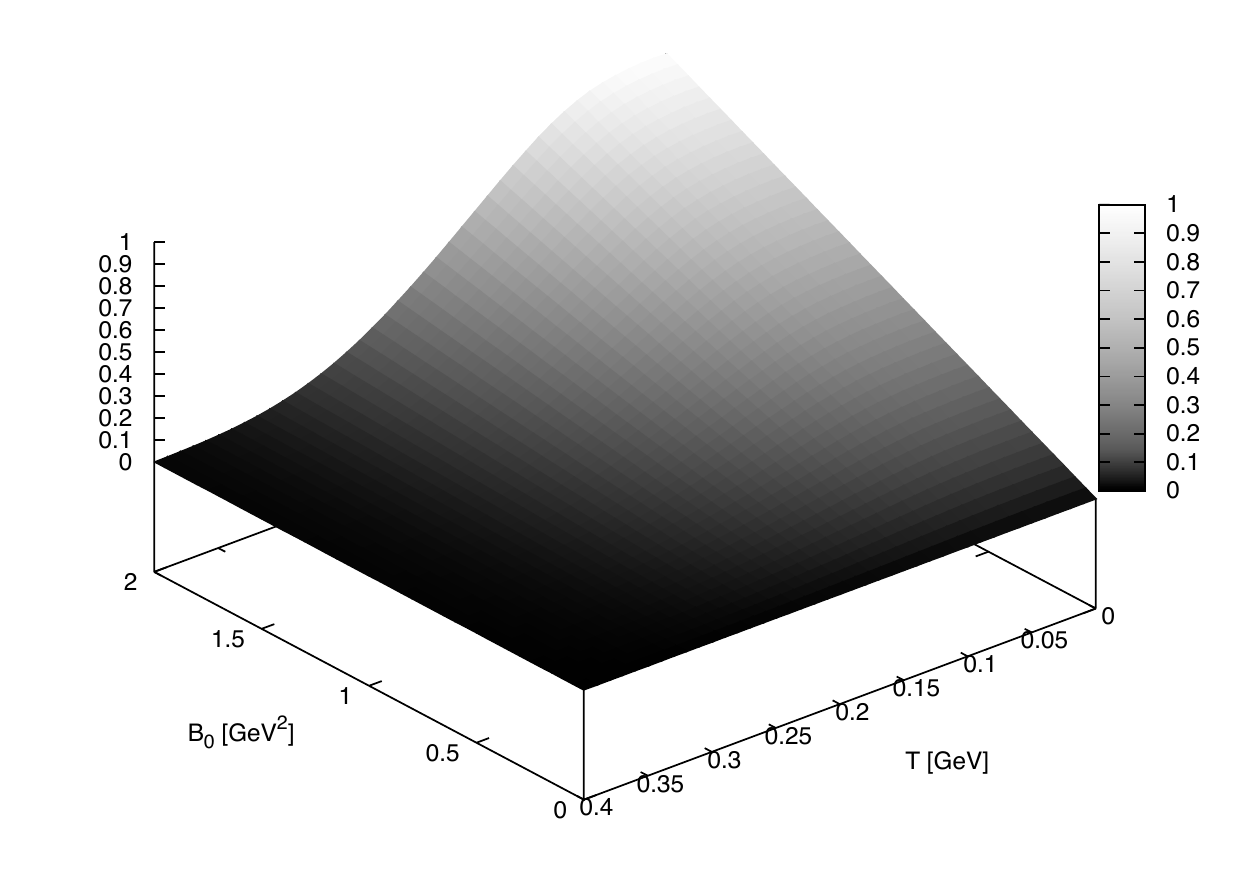}
\caption{Induced EM current indicating the CME, $|\langle j_{3,4}/(iA_{4,3})\rangle_{F,\delta}|\times10^{3}$, as functions of $B_{0}$ and $T$ for $\epsilon=10^{-3}$.}       
\label{FIG4}
\end{figure}

In the left panel of Fig.~\ref{FIG3}, we draw $|\langle j_{3,4}/(iA_{4,3})\rangle_{F,\delta}|$ as functions of $B_{0}$ for $T=(0,\,50,\,100,\,150,\,200)$ MeV, separately. The curves are increasing linearly with respect to $B_{0}$ as expected, indicating the CME. Being similar to the local chiral density, the $|\langle j_{3,4}/(iA_{4,3})\rangle_{F,\delta}|$ becomes insensitive to the magnetic field, as $T$ increases. Hence, we can conclude that the signal of the CME is weakened with respect to $T$ at a certain value of $B_{0}$. In Fig.~\ref{FIG4}, we draw the $|\langle j_{3,4}/(iA_{4,3})\rangle_{F,\delta}|$ as functions of $B_{0}$ and $T$ to see its overall behavior. Although there can be a increasing competition between the instanton and sphaleron contributions in the high-$T$ region ($T>T^{\chi}_{c}$), it turns out that, in our present framework, the instanton contribution to the CME becomes about one-tenth at $T\approx280$ MeV in comparison to its maximum and almost disappears above $T\approx300$ MeV as show in the figure.

\section{Summary and conclusion}
In the present work, we have investigated the chiral magnetic effect (CME), which signals the nontrivial topological charge $Q_{\mathrm{t}}$ of the QCD vacuum. The evidence of the CME can be explored by measuring the electromagnetic (EM) current induced by the external magnetic field, which is generated from the non-central heavy-ion collision. We were interested in the CME in the low-temperature region below the chiral phase transition, $T\lesssim T^{\chi}_{c}\approx200$ MeV. To this end, we employed the instanton vacuum configuration with the nonzero instanton-number fluctuation $\Delta$, which is proportional to the $Q_{\mathrm{t}}$. In order to incorporate with finite $T$, we made use of the Harrington-Shepard caloron with the trivial holonomy for the $T$-dependent constituent-quark mass. The fermionic Matsubara formula was taken into account for the anti-periodic sum over the Euclidean time direction. We calculated the vacuum expectation values of the local chiral  density, chiral charge density, and EM current, which are induced by the  external magnetic field with the nontrivial $Q_{\mathrm{t}}$, using the linear Schwinger method and functional derivatives of the effective action with the external sources. We expanded the relevant matrix elements up to $\mathcal{O}(\delta^{2})$, where the $\delta$ represents the strength of the $P$- and $CP$-violation and is proportional to the $Q_{\mathrm{t}}$. We have observed the followings:
\begin{itemize}
\item The transverse EM current induced by the external magnetic field is far smaller than that of the longitudinal one, $j_{\parallel}\gg  j_{\perp}$, and their ratio is proportional to $\delta$, $|j_{\parallel}/j_{\perp}|\propto\delta\sim Q_{\mathrm{t}}$~\cite{Buividovich:2009wi}.
\item We obtain a relation such that the strength of the chiral charge density is the same with that of the induced EM current in the $\hat{x}_{3}$- and $\hat{x}_{4}$-directions: $\langle n_{\chi} \rangle=|\langle j_{3,4} \rangle|$ as long as the Lorentz invariance remains unbroken~\cite{Fukushima:2008xe,DHoker:1985yb}. 
\item The correct relation between the fourth component of the EM field and  the the chiral chemical potential is found: $i\delta A_{4}=\mu_{\chi}$~\cite{Fukushima:2008xe}.
\item The CME, represented by the local chiral density and induced EM current, becomes insensitive to the external magnetic field as $T$ increases. This can be understood by that the instanton, {\it i.e.}  tunneling effect gets diminished with respect to $T$: $Q_{\mathrm{t}}\to0$~\cite{Buividovich:2009wi}.
\item The local chiral density and induced EM current as functions of $T$ decrease faster for the larger external magnetic field, since the decreasing instanton effect is strengthened. 
\item The induced EM current from the instanton contribution decreases smoothly with respect to $T$ and becomes one-tenth at $T\approx280$ MeV and negligible beyond $T\approx300$ MeV,  if we ignore the sphaleron effect. 
\end{itemize} 
To have more information on the CME and compare with the lattice simulation, we want to investigate the correlation functions of the chiral density and the induced EM current as done in Ref.~\cite{Buividovich:2009wi}. Related works are under progress and appear elsewhere. 

\section*{Acknowledgment}
The author thanks K.~Fukushima, M.~M.~Musakhanov, and C.~W.~Kao for fruitful discussions. He is also grateful to K.~S.~Choi for technical supports on the numerical calculations. This work was supported by the NSC96-2112-M033-003-MY3 from the National Science Council (NSC) of Taiwan. 

\end{document}